\documentstyle[12pt]{article}
\catcode`\@=11 
\def\beq{\begin{equation}}
\def\eeq{\end{equation}}
\def\bea{\begin{eqnarray}}
\def\eea{\end{eqnarray}}

\def\to{\rightarrow}
\def\a{\alpha} \def\b{\beta}  \def\G{\Gamma}
\def\d{\delta} \def\D{\Delta} \def\ee{\varepsilon} 
   
 \def\l{\lambda}  \def\m{\mu} \def\n{\nu}

 \def\f{\phi}   
\def\Ps{\Psi} \def\o{\omega}  
\def\pa{\partial}   \def\half{{1\over
2}} \def\dx{d^3\!x\ }  
 
\def\det{\,{\hbox{det}}\,} 
 \def\ie{{\it i.e.}} 
 \def\rhs{right hand side}

\def\NP{{\it Nucl.~Phys.~}}

\def\vol#1{{\bf #1}}
\def\vyp#1#2#3{\vol{#1} (#2) #3}

\def\A{$A_i^a$}
\def\u{$u_i^a$}

\begin{document}

\begin{flushright}
McGill/95-30 \\
{\tt hep-th/9505188}
\end{flushright}

\begin{center}
\vglue 0.50cm
{\large\bf NEW GAUGE INVARIANT VARIABLES\\ FOR YANG-MILLS
THEORY\footnote{Talk given at the {\em XVII
Montr\'eal-Rochester-Syracuse-Toronto Conference}, Rochester,
NY, May 1995. E-mail: {\tt haagense@cinelli.physics.mcgill.ca}.}\\}
\vglue 1cm
{\normalsize Peter E. Haagensen\\}
\vglue 0.7cm
{\normalsize\em Physics Department, McGill University\\
3600 University St.\\
Montr\'eal~~H3A~2T8~~~CANADA\\}
\vspace{1cm}
\end{center}

\begin{abstract}

A new set of gauge invariant variables is defined to describe the
physical Hilbert space of $d\!=\!3\!+\!1$ $SU(2)$ Yang-Mills theory in
the fixed-time canonical formalism.  A natural geometric interpretation
arises due to the $GL(3)$ covariance found to hold for the basic
equations and commutators of the theory in the canonical formalism.  We
emphasize, however, that we are not interested in and do not consider
the coupling of the theory to gravity.  We concentrate here on a
technical difficulty of this approach, the calculation of the electric
field energy.  This in turn hinges on the well-definedness of the
transformation of variables, an issue which is settled through
degenerate perturbation theory arguments.

\end{abstract}
\vspace{1cm}

\begin{flushleft}
McGill/95-30 \\
May 1995
\end{flushleft}

\vfill\eject

\setcounter{page}{1}
\pagestyle{plain}
\baselineskip=15pt

{\parindent=0em\bf \large 1. Introduction}\vspace{.3cm}

While there is no doubt nowadays that Quantum Chromodynamics is the
governing theory of the strong interactions, and that confinement and
chiral symmetry breaking are the essential mechanisms responsible for
the low-energy spectrum of mesons and baryons observed experimentally,
there is on the other hand no established and systematic procedure to
treat the theory analytically in this low-energy regime, where a
perturbative
expansion in the gauge coupling is not possible. Moreover, since the
early days of QCD there has even been a plausible explanation of what
could be responsible for the basic mechanisms of low-energy QCD, namely,
the dual superconductor picture of the vacuum and magnetic monopole
condensation. Still, as far as analytical treatment goes, no
well-defined way to proceed further has yet been determined, and one has
been mostly left with general pictures rather than specific calculations
with definitive results.

At the same time, it is generally accepted that, whatever the physical
picture may be in whatever energy regime, a correct treatment must
include proper determination of the gauge invariant degrees of freedom
and implementation of gauge invariance. This is true of the high energy
perturbative regime, and is also what the dual superconductor picture
attempts to do.

In this paper we would like to propose a systematic implementation of
this gauge invariance principle, with a view to study the low-energy
regime of QCD. In fact, our basic setting will be more
limited: if one is content with considering color confinement separately
from chiral symmetry breaking, and if one considers that it is gluons
and not quarks that are essential in generating confinement, then
one may as well study the simplified model of pure $SU(2)$ Yang-Mills
theory, and this is what we do here. Without quarks, there is no chiral
symmetry to speak of, let alone its breaking, and $SU(2)$ is not the
gauge group of QCD, and thus whatever we study will admittedly not be
the QCD vacuum and probably not even a good approximation to it.
However, we are motivated by the fact that this simplified model does
present the feature of color confinement, and a sufficiently intricate
and interesting vacuum structure, radically different from that of
perturbative or of non-gauge theories.

The general way we propose to implement gauge invariance is simple
enough to state:  in the fixed-time canonical formalism, rather than
using the vector potential \A\ as basic Hilbert space variables, we
would like to use instead variables which are invariant under gauge
transformations.  Wavefunctionals which are functions exclusively of
these variables will then automatically implement Gauss' law, \ie , will
be gauge invariant.  This is unfortunately and by far not the end of the
story:  if these variables span too small a sample of Hilbert space, if
they do not represent the correct number of gauge invariant degrees of
freedom, if the transformation of variables is ill-defined for certain
field configurations, then this way to proceed might not be feasible.

In Sec.~2, we will briefly mention one such set of variables, which is
inappropriate due to the presence of Wu-Yang ambiguities.  We then
describe a new set of gauge invariant variables which is free of these
ambiguities.  As opposed to the Wu-Yang ambiguous variables, these new
variables depend nonlocally on \A , and form an ``overcomplete" set,
much like Wilson loops.  A natural geometrical description will emerge,
without any coupling to gravity, due to the $GL(3)$ symmetry already
present in all the basic formulas and commutators of the canonical
formalism except for the Hamiltonian itself.  We exploit this geometry
in Sec.~3, where we consider the transformation of the theory to the new
variables.  In Sec.~4, for the sake of illustration, we consider a
specific class of configurations, namely, those for which the new gauge
invariant variables describe an Einstein geometry.  Geometries with a
high degree of symmetry such as this one may at first seem to present a
problem for our new variables because of some subtleties regarding the
0-modes of a certain operator.  In Sec.~5 we turn to these subtleties.
They relate in particular to the computation of the electric field
energy, a technically more involved point in this approach, and the
well-definedness of the transformation of variables. Settling these
issues is our main objective here; we will not dwell on any details of
calculation since these can be found in previous work
along these lines\cite{hj}. We also refer the interested reader to
\cite{hj} for two issues that will not be considered here, namely, the
introduction of static point color sources in this formalism, and the
extension to gauge group $SU(3)$.\vspace{.6cm}

{\parindent=0em\bf \large 2. Canonical Formalism and Gauge Invariant
Variables}\vspace{.3cm}

In the fixed-time canonical formalism in $A_0^a=0$ gauge, the
Hamiltonian is
\beq\label{1}
H\ =\ \half\ \int \dx \left({B^{ia}B^{ia}\over
g_s^2}+ g_s^2E^{ia}E^{ia}\right)\ ,\eeq
where the color magnetic field $B^{ia}$ is defined as a function of the
vector potential $A^a_i$:\footnote{The reason for the peculiar placement
of the spatial indices will become clear in what follows.}
\beq \label{2} B^{ka}[A]\
=\ \ee^{kij}F^a_{ij}[A]\ =\ \ee^{kij} (\pa_i A^a_j+\half\ \ee^{abc}A^b_i
A^c_j)\ ,\eeq and the color electric field $E^{ia}$ is the canonical
momentum conjugate to $A^a_i$, such that:
\beq\label{3}
[\ A^a_i(x),E^{jb}(y)\ ]\ =\ i\d^{ab}\d^j_i\d (x-y)\ .\eeq
In the particular realization we will use of the above commutators the
electric field is simply a functional derivative with respect to \A .
The Gauss law generator is,
\beq \label{4}
{\cal G}^a(x)\ =\ D_i E^{ia}(x)\
\equiv\ \pa_i E^{ia}(x)+ \ee^{abc}A^b_i(x)E^{ic}(x)\ ,\eeq
and it generates the following infinitesimal gauge transformations:
\bea
\left[\ A^a_i(x),{\cal G}^b(y)\ \right]\ & = & \ -i\d^{ab}\pa_i\d (x-y)\
+i\ee^{abc}A^c_i(x)\d (x-y) \label{5}\\
\left[\ E^{ia}(x),{\cal G}^b(y)\ \right]\ & = & \ i\ee^{abc}E^{ic}(x)\d
(x-y) \label{6}\\
\left[\ H,{\cal G}^a(x)\ \right]\ & = & \ 0\  ,
\label{7}\eea
with $B^{ia}$ transforming identically to $E^{ia}$.
These generators furthermore form a local algebra
\beq\label{8}
[\ {\cal G}^a(x),{\cal G}^b(y)\
]\ =\ i\ee^{abc}{\cal G}^c(x)\d (x-y)\ .\eeq

What we are looking for, then, are variables which represent the correct
number of gauge invariant degrees of freedom at each point, and which
are annihilated by ${\cal G}^a$. The simplest possibility is of course
to consider the magnetic field. As the above shows, unlike \A\ it simply
rotates under ${\cal G}^a$, and so one can easily build something gauge
invariant from that: $\f^{ij}\equiv B^{ia}B^{ja}$. Furthermore,
$\f^{ij}$ defined thus is a $3\times 3$ symmetric matrix, which has 6
independent entries, \ie, just the correct number of gauge
invariant degrees of freedom for $SU(2)$. Unfortunately, this approach
does not work, due to the fact that gauge field configurations in
general have Wu-Yang ambiguities\cite{fhjl}. These are gauge-unrelated
configurations of \A\ which lead to the same magnetic field. If
$\f^{ij}$ are used as basic variables, they are blind to Wu-Yang
ambiguous configurations and thus such configurations will not be
integrated over when they actually should. There does not seem to be any
simple way out of this problem, and thus we discard this first
possibility, and turn to one which does not suffer from these
ambiguities.

To motivate our definition of gauge invariant variables, let us note
first of all that Eqs.~(\ref{2})-(\ref{6}),(\ref{8})
are covariant under $GL(3)$
transformations if we require \A\ to transform as a covariant vector
and $E^{ia}$ as a contravariant vector density under $GL(3)$:
\bea\label{9}\label{10}
E^{'ia}(x')&=&\left|{\pa x\over\pa x'}\right| {\pa x^{'i}\over\pa x^{j}}
E^{ja}(x)\\
A^{'a}_i(x')&=& {\pa x^{j}\over\pa x^{'i}} A^a_j(x)\ ,\eea
where $x^i\to x^{'i}(x)$ is some coordinate reparametrization, and
$\pa x^{'i}/\pa x^{j}$ is a $GL(3)$ matrix. In fact, we can see
that the only failure of the canonical formalism to fulfill this
symmetry comes in the Hamiltonian, where the space indices are summed
with a Kronecker $\d_{ij}$. This symmetry will serve as a useful tool in
what follows. We now wish to define a set of new variables $u^a_i$
that will respect not only the $SU(2)$ gauge symmetry
but also this $GL(3)$ symmetry. We define them as
follows, through a system of linear first order p.d.e.'s:
\beq\label{11}
\ee^{ijk}D_ju^a_k=\ee^{ijk}(\pa_ju^a_k+\ee^{abc}A^b_j u^c_k)\ =\ 0\ .
\eeq
A few comments are in order with respect to this definition.
Firstly, as opposed to the previous gauge covariant variables
$B^{ia}[A]$, \A\ is built locally as a function of \u\ (in fact, the
explicit relation $A^a_i[u]$ is simple to find from the above), whereas
it is
\u\ that is built nonlocally as a function of \A . This manifestly
eliminates Wu-Yang ambiguities, because a single \u\ cannot give rise to
two different \A\ configurations. Secondly, it is also true on the
other hand that there will always be many configurations \u\ for a
single \A : if \u\ is a solution, then so is $\l u^a_i$ for $\l$ any
real number, and there might
even be many other solutions in special cases, as we shall see in what
follows. Finally, we see that this definition is both $SU(2)$ and
$GL(3)$ covariant if \u\ transforms as an adjoint $SU(2)$ vector and
$GL(3)$ covariant vector.

{}From the \u\ we are now ready to build our gauge invariant variables.
They are:
\beq g_{ij}=u^a_iu^a_j\ .\eeq\vspace{.6cm}

{\parindent=0em\bf \large 3. Geometry and Yang-Mills Fields}
\vspace{.3cm}

{}From Eq.~(\ref{11}), it is straightforward to show that
\beq
\pa_iu^a_j+\ee^{abc}A^b_i u^c_j -\G^k_{ij}u^a_k=0\ ,\eeq
where $\G^i_{jk}$ is the Christoffel connection built from $g_{ij}$.
A geometry has thus entered the formalism without the
introduction of any extraneous coordinate reparametrization invariance.
{}From the above, we see that \u\ is playing the role of a 3-bein,
$g_{ij}$ that of a metric, and $\o^{ac}_i=\ee^{abc}A^b_i$ that of a spin
connection.

We now list the expressions for the relevant Yang-Mills quantities
in terms of \u , where the geometrical nature of the formalism becomes
manifest. We do not present any deductions here, as they can all be
found in \cite{hj}. The magnetic field is found from the gauge Ricci
identity, expressing it in terms of the commutator of two covariant
derivatives. One finds
\beq B^{ai}\ =\ \sqrt{g}\ u_j^a\ G^{ij} \ ,\eeq
where by $\sqrt{g}$ we mean $\det u^a_i$, and $G_{ij}$
is the Einstein tensor built from the metric $g_{ij}$.
The gauge Bianchi identity reads:
\beq 0=D_i B^{ai}\ =\ \sqrt{g}\ (\nabla_i G^{ij})\ u_j^a\ ,\eeq
so that it stands in mutual implication with the geometric Bianchi
identity on the \rhs\ above.

In order to find the electric field we define the tensor $e^{ij}$:
\beq {\delta\over \delta A_i^a}\ \equiv\ \sqrt{g}\ u_j^a e^{ij}\ .\eeq
In the geometrical variables, Gauss' law becomes
\beq i {\cal G}_a \ =\ D_i \left({\delta\over \delta A_i^a} \right)=
\ \sqrt{g}\ u_j^a (\nabla_i e^{ij})\ .\eeq
Thus, if $\Psi$ is a gauge invariant wavefunctional, then
\beq\nabla_i e^{ij} \Psi\ =\ 0\ ,\eeq
and vice-versa. It is also possible to show\cite{hj} that if $\Ps$ is
only a function of $g_{ij}$ then it is gauge invariant, and that if it
is gauge invariant, then it is only a function of $g_{ij}$.

In order to calculate the electric field, we consider the response of a
wavefunctional to a small variation $\d A^a_i$ and attempt to express
that in terms of variations in the ``metric" $\d g_{ij}$. We find, for
wavefunctionals depending solely on $g_{ij}$:
\beq\label{19}
{\d\Psi\over\delta g_{ij}} =
{\ee^{mni}\over 2} \nabla_m \tilde e^j_{\ n} \Psi\ ,\eeq
where
\beq
{\tilde e}^i_{\ j}\equiv e^i_{\ j}- {1\over 2}\ \delta^i_j
e^s_{\ s}\ .\eeq
It would now remain to invert the operator $T=\ee\nabla$ on the
\rhs\ above to find the electric field in terms of metric variations.
Formally, one may consider the eigenvalue problem for this operator:
\beq
T^i_{\ n}\phi_\alpha^{nj}\equiv {1\over\sqrt{g}}\
\ee^{im}_{\ \ \ n}\nabla_m \phi_\alpha^{nj} =\lambda_\alpha
\phi_\alpha^{ij} \ .\eeq
Once one has the spectrum, inversion can be obtained through the
standard spectral representation. The presence of 0-modes of $T$,
however,
apparently leads to a problem: the inversion can only be made in the
subspace orthogonal to the 0-modes of $T$, and
it seems there
is an indetermination in the components of the electric field in the
direction of 0-modes of $T$, which would lead to spurious
constraints. We investigate this more carefully in Sec.~5. For
now, let us outline how this problem is circumvented: Eq.~(\ref{19}) is
gotten essentially through the chain rule
\beq
{\d\Psi\over\d g_{ij}} ={\d u^a_k\over\delta g_{ij}}\cdot
{\d A^b_\ell\over\d u^a_k }\cdot {\d\Psi\over\d A^b_\ell}\ .\eeq
It is the 0-modes of $\d A^b_\ell /\d u^a_k$ which cause the problem.
The chain rule in the opposite direction, to isolate $\d\Psi /
\d A^b_\ell$ would require a
well-defined expression for $\d u^a_k /\d A^b_\ell$. Strictly speaking,
however, while \A\ is a function of \u , \u\ is not a function of \A ,
since there are many \u\ which lead to the same \A . This leads to the
indetermination in the electric field mentioned above. To cure it, we
will need a
systematic prescription that assigns a definite variation $\d u^a_i$ to
a variation $\d A^a_i$. This as we will see is gotten through standard
quantum mechanical degenerate perturbation theory.

In the following section we consider the spectrum of $T$ for a special
class of configurations, those for which $g_{ij}$ describes an Einstein,
or constant curvature, space. This will in particular illustrate the
0-mode problems appearing in the inversion of $T$.
\vspace{.6cm}

{\parindent=0em\bf \large 4. Einstein Space Configurations}
\vspace{.3cm}

To have an idea of what the spectrum of $T$ looks like, we consider in
this section those configurations whose associated metric describes an
Einstein space (\ie, $R_{ij}\propto g_{ij}$, which in $d\!=\!3$ implies
a constant curvature space).  It is possible to show that if $g_{ij}$ is
Einstein with positive curvature (\ie, the sphere $S_3$), then in some
gauge $u^a_i =a^{-1}\ A^a_i$, where $a^{-1}$ is the radius of the
sphere. This in turn
implies that for these configurations, $A^a_i=\half \bar{A}^a_i$, where
$\bar{A}$ is pure gauge.  Throughout this section we consider only the
sphere $S_3$; the hyperbolic case works in a similar way.

To find the spectrum of $T$, we decompose $\f^{ij}$ as follows:
\beq
\f^{ij}=S^{ij}+ {1\over2}\ {\ee^{ijk}\over\sqrt{g}}\ V_k+ {1\over3}\
g^{ij}\bar{\f}\ ,\eeq
where $S^{ij}$ is symmetric and traceless. By trying different {\em
ans\"atze} for $S^{ij}, V_k$ and $\bar{\f}$ one then finds the following
eigenvectors and eigenvalues of
$T$:
\begin{enumerate}
\item $\f_{ij}={1\over2s_N^2}\left[ \nabla_i\nabla_j+s_N^2g_{ij}
-\sqrt{s_N^2-a^2}\ {\ee_{ijk}\over\sqrt{g}}\ \nabla^k\right] Y_{Nlm}$,
with eigenvalues $\l\!=\!\pm\sqrt{s_N^2\!-\!a^2}$, where
$-s_N^2=-a^2N(N+2), N=1,2,...$ is
the spectrum of the Laplacian acting on scalars on $S_3$ (with radius
$a^{-1}$), and
$Y_{Nlm}$ are the associated hyperspherical harmonics.
\item $\f_{ij}=\left[ \nabla_i\nabla_j+a^2g_{ij}
\right] Y_{Nlm}$, all with eigenvalue $\l=0$.
\item $\f_{ij}$ is symmetric, traceless and covariantly divergence-free,
and is an eigenvector of the Laplacian acting on rank 2 tensors. The
associated eigenvalue is $\l=\pm\sqrt{t_N^2+3a^2}$, where $-t_N^2=-s_N^2+2a^2$
is the spectrum of the Laplacian acting on rank 2 tensors on $S_3$.
\item $\f_{ij}=-{1\over2}(\l-a^2/\l)^{-1}(\nabla_iV_j+\nabla_jV_i)
+{1\over2} {\ee_{ijk}\over\sqrt{g}}\ V^k$, where $V_i$ is a covariantly
divergence-free eigenvector of the Laplacian acting on vectors. The
associated eigenvalues are $\l=\pm\left( \sqrt{v_N^2+2a^2}\pm
\sqrt{v_N^2-2a^2}\right)$, where $-v_N^2=-s_N^2+a^2$ is the spectrum of
the Laplacian acting on vectors on $S_3$.
\end{enumerate}

We believe this exhausts the spectrum of $T$ on $S_3$, but we have no
complete proof of this.  Two features of this spectrum are worth
mentioning:  first, there is an infinite set of exact 0-modes.  To this
infinite set, there corresponds an infinite set of solutions of
Eq.~(\ref{11}) (cf.~\cite{hj}).  Second, some of the eigenvalues in 4.~above
approach 0 as $1/N$ for $N\to\infty$.  Thus, there is an accumulation of
modes near $\l=0$.  Under an infinitesimal variation of \A , these
exact- and near-0 modes mix to give the variation in the solution to
Eq.(\ref{11}).  \vspace{.6cm}

{\parindent=0em\bf \large 5. Electric Energy and Degenerate
Perturbation Theory}\vspace{.3cm}

If we look at the variation of \A\ under a variation in \u ,
\beq
\d A^a_i(x) = {\d A^a_i(x)\over\d u^b_j(y)}\cdot\d u^b_j(y)\eeq
it immediately seems that if $\d A^a_i(x)/\d u^b_j(y)$ has
left zero modes (\ie, zero modes upon action of the operator to the
left), there will be constraints on $\d A^a_i(x)$: apparently not all
$\d A^a_i(x)$ can be generated from all $\d u^a_i(x)$.
Since the electric field is the response of the wavefunctional
to a variation $\d A^a_i(x)$,
this would mean in particular that, in the $u$-variables, the
electric field could not be measured along all directions. These
statements being true would not be a
good starting point for these variables. We shall now see, however, that
because of the enormous degeneracy of zero and near-zero modes,
through
degenerate perturbation theory considerations one can actually prove the
transformation of variables is well-defined, and the electric field can
be determined in all directions in the $u$-variables.

Let us reconsider Eq.~(\ref{11}) in the light of
the fact that it has many solutions for any \A , and let us label
these solutions as $u_\a$, where we omit all space and color indices;
$\a$ spans the different solutions. Picking one particular $\a\!
=\!\bar{\a}$,
one then builds a metric $g_{ij}^{\bar{\a}}$, from which one can study
the fully $GL(3)$ covariant eigenvalue problem for the nonzero modes:
\beq
{1\over\sqrt{g}}\ \ee^{ijk}D_j w^a_k=\l w^{ia}\, .\eeq
Again omitting all color and space indices, we can label these
eigenvectors by some index $a$, $w_a$, and the
corresponding eigenvalue by $\l_a$.
These states can be orthonormalized:
\bea
(u_\a,u_{\a'})&\equiv&
\int d^3\!x\ \sqrt{g}\ g^{ij} (u_\a)^a_i (u_{\a'})^a_j =
\d_{\a\a'}\\
(w_a,w_{a'})&=&\d_{aa'}\\
(u_\a,w_{a})&=&0\ .\eea
The $\d$'s on the \rhs\ may be Dirac or Kronecker
deltas, depending on whether their arguments are continuous or discrete.

We now consider ``perturbations'' $\d A^a_i(x)$. If
$\tilde{u}_\a$ denotes the perturbed $u_\a$, and $\d\l_\a$ the perturbed
eigenvalue, then:
\beq\label{pert}
\ee D\tilde{u}_\a+\D A\ \tilde{u}_\a=\d\l_\a\ \sqrt{g}\
\tilde{u}_\a\, ,\eeq
where $\D A\equiv\ee\ee\d A$ and all space and color indices are
omitted.

We can expand $\tilde{u}_\a$ in the $\{ u_\a,w_a\}$ basis:
\beq\label{exp}
\tilde{u}_\a=P_{\a\a'}\ u_{\a'} +Q_{\a a}\ w_a\, .\eeq
Here and below, repeated indices are summed over unless explicitly
indicated otherwise. The
coefficients $P$ are of order $1$, while $Q$ are of order $\d A$.
We now need solutions for $P$ and $Q$.
Inserting this in Eq.~(\ref{pert}),
and using orthonormality, we obtain the system
\bea
P_{\a\b'} (\D A)_{\b\b'}+Q_{\a a} (\D A)_{\b a}&=&
\d\l_\a\ P_{\a\b}~~~(\a,\b~ {\rm fixed})\\
P_{\a\b'} (\D A)_{b\b'}+\l_b Q_{\a b}+Q_{\a a} (\D A)_{b a}&=&
\d\l_\a\ Q_{\a b}~~~(\a,b~ {\rm fixed})\ , \eea
where $(\D A)_{\b\b'}=(u_\b,(\D A)u_{\b'})$. The last two terms
of the second equation above are of higher order and we drop them, to
find:
\beq
Q_{\a b}=-{1\over\l_b}P_{\a\b'} (\D A)_{b\b'}~~~(\a,b~ {\rm fixed}).
\eeq
This is inserted in the first equation (it is not difficult to see
that the difference of the two ``$P$-terms" is of the same order as the
``$Q$-term", and so this latter one should not be dropped):
\beq
\left[ (\D A)_{\b\b'}- (\D A)_{\b a}{1\over\l_a}(\D A)_{a\b'}
\right] P_{\a\b'}=\d\l_\a\ P_{\a\b}~~~(\a,\b~ {\rm fixed}) .\eeq
If we now diagonalize $(\D A)_{\a\b}$:
\beq
S^T_{\m\a}(\D A)_{\a\b}S_{\b\n}=(\D A)_{\m}\d_{\m\n}~~~(\m,\n~
{\rm fixed}) ,\eeq
we have in the transformed system:
\beq
\tilde{P}_{\a\n}(\D A)_{\n}- \tilde{P}_{\a\m}(\D A)_{\n
a}{1\over\l_a}(\D A)_{a\m}
=\d\l_\a\ \tilde{P}_{\a\n}~~~(\a,\n~ {\rm fixed}) ,\eeq
where $\tilde{P}_{\a\n}=P_{\a\b}S_{\b\n}$.

Since we are assuming there is a solution for every $A$-configuration,
there must be an $\a$, call it $\a=0$, such that $\d\l_\a=0$. That is
the one we are interested in. Also, there is a $\m$, call it $\m=0$,
such that $(\D A)_{\m}$ is minimum (no $(\D A)_{\m}$ can be zero, since
otherwise there would be a linear combination of $u_\a$ that would be a
solution to both $A$ and $A+\d A$ simultaneously, which we have shown
previously to be impossible). Now we can determine $P$ uniquely: the
{\em ansatz} $\tilde{P}_{0\m}=\d_{0\m}+{\cal O}(\d A)$ solves the above
to lowest order for the component $\a=0$:
\bea
\tilde{P}_{0\n}&=&{1\over (\D A)_{\n}}\ (\D A)_{\n
a}{1\over\l_a}(\D A)_{a0}~~~(\n~ {\rm fixed})\\
Q_{0b}&=&-{1\over\l_b}\ (\D A)_{b0}~~~(b~{\rm fixed}) .\eea
This is the result we need: by substituting this above in
Eq.~(\ref{exp}),
we have a prescription for the unique variation in \u , $\d u^a_i$,
given a generic variation $\d A^a_i$. This amounts to having $\d u^a_i
/\d A^b_j$, from which one can then unequivocally calculate the electric
field.\vspace{.6cm}

{\parindent=0em\bf \large 6. Conclusions}\vspace{.3cm}

It is virtually impossible to present in a short space all the details
of a long project such as this.  Many of our results, which may appear
too telegraphic here (or even do not appear here at all), can be found
in \cite{hj}, while others will be left for future publications.  Here
our main concern was with establishing the fact that the transformation
of variables and the electric field in the variables \u\ are
well-defined even for those geometries which, because of a high degree
of symmetry, lead to 0-modes of the operator $T$.  Beyond this fact, it
remains to be seen whether the technical difficulties associated to the
calculation of the electric field can be overcome at a practical level,
in order to allow for a numerical treatment of some sort.  Work along
these lines is in progress.  \vspace{.6cm}

{\parindent=0em\bf \large Acknowledgments}\vspace{.3cm}

The work presented here is being carried out in collaboration with K.
Johnson, C.S.  Lam, and R. Khuri.  It is a pleasure to thank them for
sharing their insight with me.  I would also like to thank N. Kaloper
for discussions on degenerate perturbation theory.

\end{document}